\documentclass[sigconf]{acmart}
\usepackage[normalem]{ulem}

\AtBeginDocument{%
  }

\acmConference{}{}{}{}
\acmBooktitle{}
\usepackage{lineno}

\usepackage{times}
\usepackage{latexsym}
\usepackage[T1]{fontenc}
\usepackage[utf8]{inputenc}
\usepackage{microtype}

\usepackage{bm}
\usepackage{url}
\usepackage{graphicx}  
\usepackage{float}
\usepackage{xcolor,colortbl}
\usepackage{adjustbox}
\usepackage{subfigure}
\usepackage{booktabs,multirow}
\usepackage{bigstrut}
\usepackage{paralist}
\usepackage{makecell}
\usepackage{nicefrac}       
\usepackage{microtype} 
\usepackage{epsfig}
\usepackage{diagbox}
\usepackage{framed}
\usepackage{empheq}
\usepackage{array}
\usepackage{enumitem}
\usepackage{mdwlist}
\usepackage{xspace,mfirstuc,tabulary}
\usepackage{tcolorbox}
\usepackage{algorithm}
\usepackage{placeins}
\usepackage{algpseudocode}
\usepackage{microtype}

\usepackage[normalem]{ulem}
\useunder{\uline}{\ul}{}
\usepackage{threeparttable}
\usepackage{makecell}
\usepackage{booktabs}

\usepackage{xspace}
\usepackage{natbib}
\usepackage{hyperref}
\usepackage{url}

\usepackage{graphicx}
\usepackage{subfigure}
\usepackage{multirow}
\usepackage{multicol}
\usepackage{pifont}
\usepackage{enumitem}
\usepackage{amsmath}
\usepackage{appendix}
\usepackage[table]{xcolor}
\usepackage{colortbl}
\usepackage{graphicx}

\usepackage{tikz}

\usepackage{amsmath}
\usepackage{mathtools}
\usepackage{amsthm}
\usepackage{amsfonts}

\newcommand{\para}[1]{\paragraph{\textnormal{\textbf{#1}:}}}
\newcommand{\uls}{\begin{itemize}[leftmargin=*]}
\newcommand{\ule}{\end{itemize}}
\newcommand{\ols}{\begin{enumerate}[leftmargin=*]}
\newcommand{\ole}{\end{enumerate}}
\newcommand{\li}{\item}

\title{An Evaluation of Context Length Extrapolation in Long Code via Positional Embeddings and Efficient Attention}
\keywords{Long Code Completion, Length Extrapolation, Large Language Model}

\begin{document}
\author{Madhusudan Ghosh}
\email{madhusuda.iacs@gmail.com}
\affiliation{%
  \institution{Indian Association for the Cultivation of Science}
  \city{Kolkata}
  \country{India}
}

\author{Rishabh Gupta}
\email{gupta.rishabh@in.bosch.com}
\affiliation{%
  \institution{Bosch Research and Technology Centre, Bangalore, India}
  \city{Bangalore}
  \country{India}
  }

\begin{abstract}

The rapid advancement of large language models (LLMs) has led to a significant increase in automated tools in the software engineering, capable of performing various code-related tasks such as code generation, completion, and translation. Despite these advancements, its effectiveness is constrained by fixed context lengths, limiting its ability to generalize across long, domain-specific code sequences. To address this challenge, we investigate zero-shot, inference-only methods aimed at improving position encodings and optimizing attention mechanisms. Our goal is to provide a thorough analysis of current approaches that facilitate context length extrapolation in code, particularly in the context of long code completion tasks.

\end{abstract}
\maketitle

\section{Introduction}
\label{sec:intro}
The rise of large language models (LLMs) like LLaMA-2~\cite{genai2023llama}, Codestral~\cite{mistral2024codestral}, and CodeLLaMA~\cite{roziere2023code} has significantly changed the landscape of software engineering tasks with a considerable dependence on them for the development of various automated tools. More specifically, there is a noticeable trend towards using LLMs to assist developers in various programming activities like source code generation~\cite{liu2024your} and completion~\cite{ding2024crosscodeeval}, code translation~\cite{pan2024lost}, and legacy code understanding and explanation~\cite{bhattacharya2024selective, bhattacharya2023exploring}.

Transformer-based LLMs are potent tools for understanding and generating source code. However, they are pre-trained with fixed context lengths, which limits their ability to perform length extrapolation during inference~\cite{press_train_2022, sun2022length}. This constraint poses a particular challenge when working with lengthy, domain-specific codebases, necessitating the effective extrapolation strategies.

In the literature, there are approaches to address the challenge of handling long sequences in LLMs via task-specific fine-tuning and pre-training on extensive datasets~\cite{longcontextscale, PI, CLEX, yarn, wang2024resonance}. However, these methods are resource-intensive and risk over-fitting, which can lead to decreased performance on shorter sequences. In contrast, training-free methods that employ windowed attention mechanisms~\cite{attnsink, llminf, longnet}, have been proposed. These methods often rely on local information and may overlook long-range dependencies. An alternative cost-effective approach for addressing this challenge is prompt compression which aims to preserve essential information by shortening prompts ~\cite{jiang-etal-2023-llmlingua, jiang2023longllmlingua, li-etal-2023-compressing}.
However, the suitability of these approaches might be very limited for code datasets, which are characterized by inherent dependencies between code data points for logical understanding. This makes prompt compression an unsuitable solution for code datasets.\\


In addition, a set of cost-effective, inference-only methods include hardware-aware techniques to mitigate these challenges.  \citet{dao2022flashattention} introduced the Flash Attention mechanism, which parallelizes the sequential processing of long documents by dividing the sequence into smaller blocks. Another approach proposed by \citet{kwon2023efficient} addresses the memory-intensive caching mechanism of key-value pairs for long documents handling during auto-regressive decoding. However, all these inference-only methods have primarily been investigated on plain long text documents.\\

This raises a pressing and pragmatic research question: \textit{Do training-free, inference-only methods possess the capability to adequately handle long code sequences, which inherently require a deep understanding of syntactical and hierarchical structures?} 
Thus, we perform a comparative analysis of inference-only approaches, focusing on positional extrapolation and efficient attention mechanisms. These approaches seek to tackle the challenges of context length extrapolation in long source codes, where structural information is crucial for LLMs.

\para{Our Contributions}

To summarize, the following are the main contributions in this work.

\begin{enumerate}[leftmargin=*]

\item We present a comparative analysis of inference-only methods, focusing on positional extrapolation and efficient attention mechanisms to study context length extrapolation in LLMs using long code completion tasks.\\


\item  The empirical results demonstrate that how much these techniques are effective in handling long code sequences while preserving essential syntactical and hierarchical information which is crucial for code understanding.

\end{enumerate}





\section{Related Work}
This study primarily focuses on length extrapolation in transformer-based language models applied to source code datasets. We categorize our study into two distinct classes of the approaches to address the context length extrapolation problem:

\para{Positional Encoding Based Methods}
Length extrapolation in transformers is closely tied to word positions. \citet{vaswani2017attention} introduced sinusoidal positional encodings (PEs) to handle this, a method widely used for enhancing length extrapolation \citep{neishi_relation_2019,press_train_2022,ruoss_randomized_2023}. Improving position encoding is thus crucial for enabling transformers to handle long documents. Later trainable PEs \cite{chen-etal-2021-simple}, relative PEs \cite{shaw-etal-2018-self}, Rotary PEs (RoPE) \cite{su2023roformer}, ALiBi~\cite{press_train_2022}, xPOS~\cite{sun2022length}, and CLEX~\cite{CLEX}, propose novel approaches for positional representation. Additionally, techniques like dynamic scaling, NTK-aware scaling~\cite{tancik2020fourier}, and ReRoPE~\citep{rerope2023} further extend the model's capacity for length extrapolation.

\para{Efficient Attention Based Technique}
LLMs are intrinsically restricted by narrow context windows, which hinders their ability to effectively incorporate or leverage the complete information present in lengthy sequences. Training-free attention sinks~\cite{attnsink}, a framework designed to handle very long sequences by maintaining key tokens throughout processing, thus preventing the model from losing essential information. 
\citet{winata2020lightweight} transformed each of the three matrices of self-attention mechanism into smaller matrices. Unlike traditional methods such as singular value decomposition,~\citet{saragadam2022deeptensor} uses a DNN to learn an optimal regularizer for tensor decomposition when the distributions of the tensor are non-Gaussian. 



The generation of long sequential tokens results in a workload that is bound by memory, consequently restricting the utilization of GPUs and overall throughput. To overcome this limitation, PagedAttention~\cite{kwon2023efficient} utilizes a strategy inspired by virtual memory, which involves segmenting $\mathbf{KV}$ caches into blocks to tackle issues related to internal and external fragmentation problem while autoregressive generation. On the other hand, Flash-Decoding~\cite{dao2023flash} enhances the acceleration of attention by introducing parallelization in the sequence length of keys and values, resulting in a significant improvement of up to $8\times$ in generation speed for long sequences.

The primary distinction is that current inference-only methods for context length extrapolation concentrate on plain text. In contrast, we perform a thorough comparative analysis using long code completion tasks, with the goal of identifying a generalized context length extrapolation technique applicable to long source code datasets.

\section{Context Length Extrapolation for Code}
\begin{figure*}[t]
\centering
\includegraphics[width=.95\textwidth]{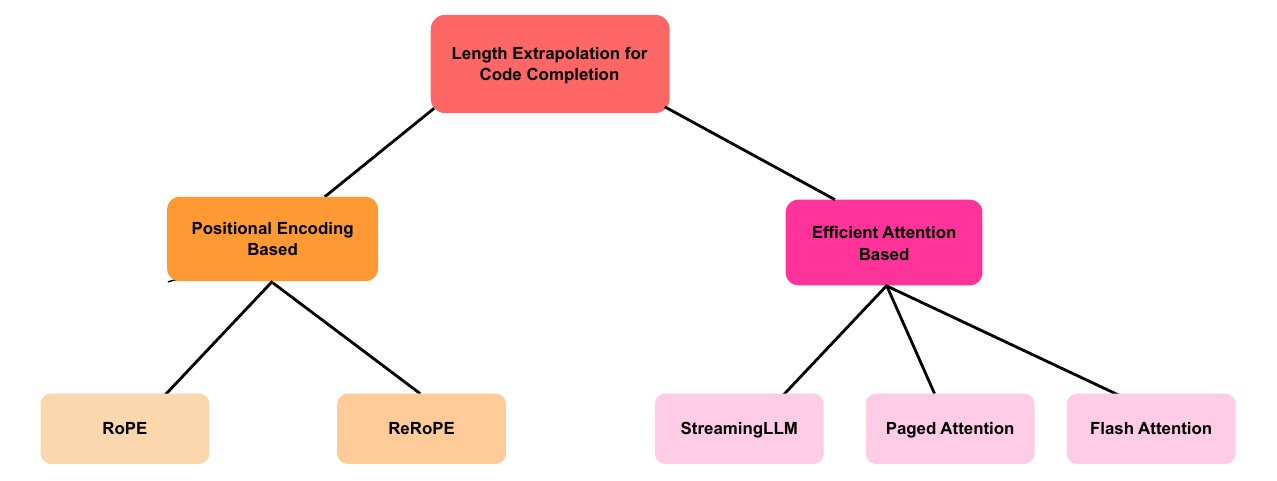}
\caption{Comparison of length extrapolation techniques for code completion, categorized into Positional Encoding Based (e.g., RoPE, ReRoPE) and Efficient Attention Based methods (e.g., StreamingLLM, Paged Attention, Flash Attention). These approaches address the challenges of handling long code sequences in Transformer models.}
\label{fig:retraining_pic}
\end{figure*}

In this section, we define the problem of context extrapolation for long code within the framework of a long code completion task, framing it as a masked token prediction challenge. 
In addition, we present the detailed explanations of various methods that we investigate for addressing the context length length extrapolation through the long code completion task.

\para{Problem Definition}
Given an incomplete source code sequence \( S = \{s_1, s_2, \ldots, s_t\} \), where \( s_i \) represents the \( i \)-th token of the code, the objective is to generate the subsequent tokens \( \hat{S}_{t+1:t+k} \) that complete the next line of the code. The generation of these tokens is conditioned on the previous context. Mathematically it can be formulated as follows:
\begin{equation}
    \hat{S}_{t+1:t+k} = \arg\max_{S_{t+1:t+k}} P_\theta(S_{t+1:t+k} \mid S)
\end{equation}

\noindent where \( P_\theta(S_{t+1:t+k} \mid S) \) denotes the posterior probability distribution over the possible sequences \( S_{t+1:t+k} \) given the context \( S \). The goal is to find the sequence of tokens that maximizes this posterior probability, thereby predicting the most likely continuation of the code sequence.

We categorize the length extrapolation methods into two different categories i.e., positional encoding based methods and effective attention based methods which can be useful to handle the long code data as depicted in the Figure~\ref{fig:retraining_pic}. 



\subsection{Positional Encoding Based Methods}

The sinusoidal positional encoding is a crucial component in transformer based models, enabling them to capture the order of input tokens~\cite{vaswani2017attention}. However, the traditional positional encoding scheme produces a constant periodic structure that fails to generalize beyond the sequence lengths observed during training, resulting in suboptimal performance when extrapolating~\cite{chi-etal-2024-attention}.


To alleviate this, relative positional encoding was introduced, which mainly incorporates the positions by capturing the relative distances between tokens rather than their absolute positions~\cite{shaw-etal-2018-self}. Although this approach helps to improve the transformer model to generalize its performance on unseen data instances by focusing on the relative distances. But even this approach cannot capture the long range dependencies while dealing with long data instances due to its primary dependencies on local context~\cite{chen2021permuteformer}.

\para{Rotary Positional Encoding (RoPE)}
In contrast to sinusoidal and relative positional encodings, \citet{su2024roformer} proposed Rotary Positional Encoding (RoPE) to overcome their limitations. RoPE applies a rotation matrix to token embeddings, capturing both relative and absolute positional information. Formally speaking, for a token \( \bm{x}_i \) at position \( i \), RoPE can be defined as follows:

\begin{equation}
\bm{x}_i' = \text{RoPE}(\bm{x}_i, i) = \bm{x}_i \cdot \bm{R}(\theta_i),
\end{equation}

where \( \bm{R}(\theta_i) \) is a rotation matrix dependent on the position \( i \) and a learnable parameter \( \theta_i \). The rotation matrix \( \bm{R}(\theta_i) \) can be formulated as follows:

\begin{equation}
\bm{R}(\theta_i) = 
\begin{pmatrix}
\cos(\theta_i) & -\sin(\theta_i) \\
\sin(\theta_i) & \cos(\theta_i)
\end{pmatrix}.
\end{equation}

In contrast to sinusoidal and relative positional encodings, RoPE effectively captures both short- and long-range dependencies by preserving the angular information of token positions. This characteristic is thought to enhance the model's ability to extrapolate context lengths.
 
\para{Rectified Rotary Positional Encoding (ReRoPE)}
RoPE outperforms the traditional positional encoding approaches by capturing both relative and absolute positional information through a rotation matrix. However, 
after a certain length static $\mathbf{R}(\theta)$ cannot scale well to encode the positional information~\cite{emozillareddit}. To address this, \citet{rerope2023} proposed an inference-only solution (ReRoPE) that leverages the pre-trained knowledge of LLMs.
ReRoPE introduces a post-processing optimization technique to extend the capabilities of RoPE by employing an attention scores based sliding window mechanism. This approach balances extrapolation of position encodings, enabling the model to handle much longer sequences smoothly.

More specifically speaking, for token positions \(i\) and \(j\) within the sliding window \(w\) (i.e., \( |i - j| < w \)), the attention score \( \alpha_{i,j}^{\text{ReRoPE}} \) is computed as follows:
\begin{equation}
\alpha_{i,j}^{\text{ReRoPE}} = \left(\mathbf{R}(\theta_i) \mathbf{q}_i\right)^\top \left(\mathbf{R}(\theta_j) \mathbf{k}_j\right),
\end{equation}

\noindent where \( \mathbf{R}(\theta_i) \) and \( \mathbf{R}(\theta_j) \) are the rotation matrices corresponding to positions \(i\) and \(j\), respectively.
For token positions \(i\) and \(j\) outside the sliding window \(w\) (i.e., \( |i - j| \geq w \)), the attention score \( \alpha_{i,j}^{\text{LeakyReRoPE}} \) is computed differently, using a scaling factor \(k\) to adjust the increase distance:

\begin{equation}
\alpha_{i,j}^{\text{LeakyReRoPE}} = \left(\mathbf{R}(\theta_i) \mathbf{q}_i\right)^\top \left(\mathbf{R}\left(\frac{\theta_j}{k}\right) \mathbf{k}_j\right),
\end{equation}

where \( \mathbf{R}\left(\frac{\theta_j}{k}\right) \) represents the scaled rotation matrix for the position \( j \). The scaling factor \( k \) effectively reduces the distance lies outside the window, ensuring that the attention mechanism can extrapolate effectively.
The final attention score \( \alpha_{i,j} \) is then computed by combining the scores from within and outside the window:

\begin{equation}
\alpha_{i,j} = \begin{cases} 
\alpha_{i,j}^{\text{ReRoPE}}, & \text{if } |i - j| < w, \\
\alpha_{i,j}^{\text{LeakyReRoPE}}, & \text{if } |i - j| \geq w.
\end{cases}
\end{equation}

\subsection{Efficient Attention Based Technique}
Although RoPE and ReRoPE mitigate some limitations of traditional positional encodings, they do not completely overcome the challenges posed by extreme length extrapolation scenarios. In such cases, effective attention-based approaches could prove beneficial.

\para{Streaming LLM}

The main concept focuses on overcoming the limitations of traditional attention mechanisms through the introduction of two key components: \textit{attention sinks} and \textit{Rolling Key-Value} (KV) Cache. In the StreamingLLM~\cite{attnsink}, the key-value cache is segmented into two parts: a rolling KV cache that retains the most recent tokens, and attention sinks that preserve the KV pairs of the initial tokens. This innovative strategy effectively maintains consistent performance while processing long documents by training within a finite attention window, enabling it to handle contexts of infinite length.

Let \( \mathbf{K}_{\text{sink}} \) and \( \mathbf{V}_{\text{sink}} \) denote the KV pairs for the attention sinks, and \( \mathbf{K}_{\text{roll}} \) and \( \mathbf{V}_{\text{roll}} \) denote the KV pairs in the rolling cache. The modified attention (StrmAttn) computation is as follows:
\begin{equation}
\begin{aligned}
\text{StrmAttn} &= \text{Softmax}\left(\frac{\mathbf{Q}_t \left[\mathbf{K}_{\text{sink}}, \mathbf{K}_{\text{roll}}\right]^\top}{\sqrt{d_k}}\right) \\
&\quad \times \left[\mathbf{V}_{\text{sink}}, \mathbf{V}_{\text{roll}}\right],
\end{aligned}
\end{equation}

\noindent where \( \left[\mathbf{K}_{\text{sink}}, \mathbf{K}_{\text{roll}}\right] \) and \( \left[\mathbf{V}_{\text{sink}}, \mathbf{V}_{\text{roll}}\right] \) represent the concatenation of the KV pairs from the attention sinks and the rolling cache, respectively.

\para{Paged Attention}
PagedAttention~\cite{kwon2023efficient} is designed to address the limitations of traditional attention mechanisms, particularly in managing memory efficiently while processing long sequences in LLMs. In general, traditional self-attention mechanism stores the key-value pair (KV) in a contiguous memory block, which leads to a very problem of effictive memory management such as internal and external fragmentation. To alleviate this challenge, Paged Attention propose a novel blockwise KV caching mechanism to store KV pair in non-contiguous fashion. Let \( \mathbf{K}_{j} = (\mathbf{k}_{(j-1)B+1}, \ldots, \mathbf{k}_{jB}) \) and \( \mathbf{V}_{j} = (\mathbf{v}_{(j-1)B+1}, \ldots, \mathbf{v}_{jB}) \) represent the key and value blocks, respectively. The attention computation in Paged Attention is performed in a block-wise manner as follows:

\begin{equation}
A_{ij} = \frac{\exp(\mathbf{q}_i^\top \mathbf{K}_j / \sqrt{d_k})}{\sum_{t=1}^{\lceil i/B \rceil} \exp(\mathbf{q}_i^\top \mathbf{K}_t / \sqrt{d_k})},
\end{equation}

\noindent where \( A_{ij} = (a_{i,(j-1)B+1}, \ldots, a_{i,jB}) \) represents the attention scores for the \(j\)-th KV block. The final attention output \( \mathbf{o}_i \) is then computed as:

\begin{equation}
\mathbf{o}_i = \sum_{j=1}^{\lceil i/B \rceil} \mathbf{V}_j A_{ij}^\top.
\end{equation}

By storing KV blocks in a non-contiguous fashion, Paged Attention allows the system to dynamically allocate memory as per requirement, reducing memory waste and improving the model's ability to perform extrapolation.

\para{Flash Attention}
Flash Attention optimizes both memory and computational complexity in attention mechanisms, especially for long sequences, by minimizing memory accesses to GPU High Bandwidth Memory (HBM) using a tiling strategy and recomputation. Flash Attention partitions input matrices \(\mathbf{Q}, \mathbf{K}, \mathbf{V} \in \mathbb{R}^{N \times d}\) into smaller blocks, loaded into faster SRAM. The attention scores for each block are computed as follows:
\begin{equation}
\mathbf{S}_{i,j} = \mathbf{Q}_i \mathbf{K}_j^\top,
\end{equation}

\noindent where \(\mathbf{S}_{i,j}\) is the score matrix for blocks \(i\) and \(j\). The softmax is then computed:
\begin{equation}
\mathbf{P}_{i,j} = \text{softmax}(\mathbf{S}_{i,j}),
\end{equation}

\noindent and the final attention output is:
\begin{equation}
\mathbf{O}_i = \sum_{j} \mathbf{P}_{i,j} \mathbf{V}_j.
\end{equation}
By keeping only small blocks in SRAM and avoiding large intermediate matrices, FlashAttention handles longer sequences efficiently.


\section{Experimental Setup}
In this section, we provide an overview of the dataset employed in our experiments. We then elaborate the various RoPE-based LLMs investigated, along with their specific configurations, to assess their capacity in processing long code sequences.

\subsection{Dataset Description}
The dataset used for this study, derived from~\citet{guo2023longcoder}, provides a comprehensive benchmark for evaluating the ability of models to handle long code sequences across three programming languages: Python, Csharp, and Java. 
Long code completion entails generating missing or incomplete segments of source code based on provided context, making it a significant challenge in software engineering. This task is complex due to the necessity of ensuring both syntactical correctness and semantic coherence over extensive contexts that can encompass hundreds or even thousands of tokens. The difficulty is further heightened in programming languages with rigid syntax and hierarchical structures, such as Csharp and Java, compared to more flexible languages like Python.\\


The dataset statistics, summarized in Table~\ref{tab:dataset_statistics}, provide insights into the average sequence lengths and their quartile distributions. Python exhibits the longest average sequence length at 3158 tokens, followed closely by Csharp at 3101 tokens and Java at 3057 tokens. All three languages share a 25\% quartile value of 3000 tokens, indicating that the shortest 25\% of sequences are of comparable lengths. However, variances arise in the median (50\%) and 75\% quartiles. Python's median sequence length is slightly longer at 3207 tokens, compared to Csharp's 3189 tokens and Java's 3134 tokens. The 75\% quartile values further highlight this trend, with Python at 3802 tokens, Csharp at 3715 tokens, and Java at 3632 tokens. These findings suggest that while handling long sequences presents challenges for all three languages, Python's longer sequences may necessitate improved extrapolation capabilities to preserve syntactical and structural integrity.

\begin{table}[t]
\centering

\begin{adjustbox}{max width=\columnwidth}
\begin{tabular}{@{}lcccc@{}}
\toprule
\textbf{Language} & \textbf{Average Length} & \textbf{25\% Quartile} & \textbf{50\% Quartile} & \textbf{75\% Quartile} \\
\midrule
\textbf{Python} & 3158 & 3000 & 3207 & 3802 \\
\textbf{Csharp} & 3101 & 3000 & 3189 & 3715 \\
\textbf{Java} & 3057 & 3000 & 3134 & 3632 \\
\bottomrule
\end{tabular}
\end{adjustbox}
\vspace{0.2cm}
\caption{Dataset statistics for code completion tasks across three programming languages: Python, Csharp, and Java. The ``25\% Quartile" indicates that 25\% of the code sequences in the dataset have a length shorter than the reported value. The ``50\% Quartile," also known as the median, is the midpoint, where half of the code sequences are shorter, and half are longer. The ``75\% Quartile" marks the point below which 75\% of the code sequences fall, representing the upper quartile.}
\label{tab:dataset_statistics}
\end{table}

\subsection{Large Language Models (LLMs)}
Similar to \citet{zhang2024hirope}, we have considered RoPE-based LLMs such as LLaMA-2 (7B)~\cite{llama2}, Sheared-LLaMA (1.3B)~\cite{shearedllama}, TinyLlama (1.1B)~\cite{tinyllama},  Vicuna (7B)~\cite{chiang2023vicuna} to conduct zero-shot inference in a low-resource scenario. 
The detailed model statistics is presented in Table~\ref{tab:model_stat}.

\begin{table}[!t] 
\footnotesize
    \centering
    \resizebox{\columnwidth}{!}{
\centering
\begin{tabular}{l|cccc}
\toprule
               & \textbf{LLaMA-2} & \textbf{ShearedLLaMA} & \textbf{TinyLLaMA} & \textbf{Vicuna} \\
\midrule
\textbf{Para.}          & 7B      & 1.3B         & 1.1B      & 7B     \\
\midrule
$\mathbf{L}_{\textbf{pretrain}}$   & 4096    & 4096         & 2048      & 2048   \\
\midrule
\textbf{Vocab Size}     & 32000   & 32000        & 32000     & 32000  \\
\midrule
\textbf{Hidden Size}    & 4096    & 2048         & 2048      & 4096   \\
\bottomrule
\end{tabular}

}
\vspace{0.2cm}
\caption{Model statistics for various investigated Large Language Models (LLMs) used in code completion tasks. The table compares LLaMA-2, ShearedLLaMA, TinyLLaMA, and Vicuna across several parameters, including the number of parameters (Para.), the maximum pretraining sequence length ($L_{pretrain}$), vocabulary size, and hidden size.}
   \label{tab:model_stat}
\end{table}

\subsection{Parameter Settings}
For ReRoPE experiment, to apply the sliding window mechanism in long code completion task, we set the window size ($w = 512$). We keep rest of the hyper parameters as mentioned in the corresponding works. Additionally, we use greedy decoding strategy to generate next 100 tokens for the long code completion task.

\subsection{Evaluation Metrics}
The evaluation of models in this study is based on two key metrics: \textbf{Exact Match (EM)} and \textbf{Edit Similarity (Edit Sim)}, as outlined by~\citet{guo2023longcoder}. The EM metric calculates the percentage of predictions that perfectly align with the ground truth sequence. This serves as a stringent evaluation criterion, where even small discrepancies, such as the omission of a single character or token, lead to a mismatch. While the EM metric is instrumental in measuring a model's capacity to reproduce precise sequences, it can be overly harsh, penalizing instances where minor inaccuracies do not hinder the overall functionality or logic of the code.

The Edit Sim, in contrast, examines the structural similarity between the predicted sequence and the ground truth, considering slight variations such as formatting changes or equivalent expressions. This metric offers a more nuanced perspective on the model's performance by emphasizing its capacity to maintain semantic and syntactical coherence, even in the absence of exact matches.


\section{Results and Analysis}
\begin{table*}[t]
\centering
\small
\begin{tabular}{@{}l@{~~~}l cc| cc| cc@{}}
\toprule
\multirow{2}{*}{\textbf{Model}} & \multirow{2}{*}{} 
& \multicolumn{2}{c}{\textbf{Python}}  & \multicolumn{2}{c}{\textbf{Csharp}} & \multicolumn{2}{c}{\textbf{Java}} \\
\cmidrule(r){3-4} \cmidrule(r){5-6} \cmidrule(r){7-8}
& & \textbf{EM} & \textbf{Edit Sim} & \textbf{EM} & \textbf{Edit Sim} & \textbf{EM} & \textbf{Edit Sim} \\
\midrule

\multirow{5}{*}{\textbf{TinyLlama}} 
& \textbf{RoPE} & 0.000 & 8.886 & 0.025 & 10.533 & 0.000 & 11.141 \\
& \textbf{ReRoPE} & 0.000 & \textbf{19.271} & 0.000 & \textbf{20.241} & 0.012 & \textbf{18.821} \\
& \textbf{StreamingLLM} & 0.000 & 12.656 & 0.000 & 11.769 & 0.000 & 11.693 \\
& \textbf{Paged Attn} & \textbf{0.034} & 2.278 & \textbf{0.087} & 4.194 & \textbf{0.024} & 3.019 \\
& \textbf{Flash Attn} & 0.000 & 11.031 & 0.000 & 11.629 & 0.000 & 11.672 \\
\midrule

\multirow{5}{*}{\textbf{Vicuna}} 
& \textbf{RoPE} & 0.013 & 23.941 & 0.000 & 15.386 & 0.000 & 15.128 \\
& \textbf{ReRoPE} & 0.000 & \textbf{24.630} & 0.000 & 23.189 & 0.000 & 21.145 \\
& \textbf{StreamingLLM} & 0.000 & 18.925 & 0.000 & 15.428 & 0.000 & 15.006 \\
& \textbf{Paged Attn} & \textbf{0.377} & 22.752 & \textbf{0.851} & \textbf{25.178} & \textbf{0.779} & \textbf{24.378} \\
& \textbf{Flash Attn} & 0.013 & 23.919 & 0.000 & 25.021 & 0.000 & 23.553 \\
\midrule

\multirow{5}{*}{\textbf{Sheared-LLaMA}} 
& \textbf{RoPE} & \textbf{0.013} & 17.287 & 0.000 & 17.286 & 0.000 & 16.734 \\
& \textbf{ReRoPE} & 0.000 & \textbf{18.144} & 0.000 & \textbf{18.515} & 0.000 & \textbf{17.660} \\
& \textbf{StreamingLLM} & 0.000 & 17.602 & 0.000 & 15.434 & \textbf{0.012} & 16.022 \\
& \textbf{Paged Attn} & \textbf{0.013} & 15.129 & \textbf{0.013} & 16.251 & \textbf{0.012} & 15.759 \\
& \textbf{Flash Attn} & \textbf{0.013} & 17.302 & 0.000 & 17.250 & 0.000 & 16.687 \\
\midrule

\multirow{5}{*}{\textbf{Llama 2}} 
& \textbf{RoPE} & 0.000 & 12.204 & 0.012 & 13.754 & 0.012 & 12.796 \\
& \textbf{ReRoPE} & 0.000 & \textbf{22.848} & 0.000 & \textbf{24.957} & 0.000 & \textbf{21.815} \\
& \textbf{StreamingLLM} & 0.000 & 18.538 & 0.000 & 16.150 & 0.012 & 15.581 \\
& \textbf{Paged Attn} & \textbf{0.104} & 19.912 & \textbf{0.038} & 22.497 & \textbf{0.024} & 21.168 \\
& \textbf{Flash Attn} & 0.013 & 12.215 & 0.013 & 13.791 & 0.012 &12.750 \\
\bottomrule
\end{tabular}
\vspace{0.2cm}
\caption{A Comparison of different models and their variants on code completion tasks across three programming languages: Python, Csharp, and Java. Each model's performance is evaluated using two metrics: Exact Match (EM) and Edit Similarity (Edit Sim). EM represents the percentage of code completions that exactly match the ground truth, while Edit Sim measures the similarity between the predicted and actual code, accounting for minor variations. It is to be noted that the highest scores for each metric across all model variants for a particular language are highlighted in bold. Bold-faced values indicate the best-performing variants for a specific metric and programming language.}
\label{tab:result_table}
\end{table*}

\label{ss:rqs}
As outlined in Section~\ref{sec:intro}, length extrapolation for code datasets presents a significant challenge compared to plain text datasets. Furthermore, the existing literature predominantly focuses on length extrapolation from the perspective of positional encoding. Therefore, this work aims to explore the following research questions (RQ).
\uls
\li \textbf{RQ-1}: \textbf{Impact of Efficient Attention Mechanisms:}
How efficient attention mechanisms impact the performance of LLMs in code length extrapolation across programming languages?
\ule

\uls
\li \textbf{RQ-2}: \textbf{Positional Extrapolation vs Efficient Attention:}
How does positional extrapolation (e.g., RoPE, ReRoPE) in LLMs performs in the code length extrapolation scenarios 
when compared with the efficient attention mechanisms?
\ule

\uls
\li \textbf{RQ-3}: \textbf{Code Language-Specific Performance:}
How does the syntax and structure of different programming languages (Python, Csharp, Java) affect the performance of LLMs for code completion task in length extrapolation scenarios?
\ule

\para{Impact of Efficient Attention Mechanisms} In relation to RQ-1, Table~\ref{tab:result_table} indicates that the efficient attention mechanism, Paged Attention, generally surpasses positional extrapolation-based methods (RoPE and ReRoPE) in terms of EM scores across all programming languages. For instance, in a zero-shot code completion scenario with Python, Paged Attention attains an EM score of $0.377$, which is significantly higher than RoPE's $0.013$. Similar patterns are evident in Csharp and Java, where Paged Attention consistently records the highest EM scores, highlighting its effectiveness in exact match situations.


Paged Attention frequently excels in EM scores but significantly lags behind in Edit Sim scores. For example, when applied to the LLaMa2 model with Python, Paged Attention attains an EM score of $0.104$ but only manages an Edit Sim score of $19.912$, markedly lower than ReRoPE’s $22.848$. This indicates that while Paged Attention is effective at generating exact matches, it struggles to preserve the overall structure and similarity of the code, resulting in reduced Edit Sim scores. This discrepancy may arise from Paged Attention's design, which prioritizes efficiency and speed, potentially overlooking less critical aspects of the input sequence that are essential for maintaining the structural integrity of the code data.

\para{Positional Extrapolation vs. Efficient
Attention} 
In reference to RQ-2, it is evident from Table~\ref{tab:result_table} that RoPE-based positional extrapolation methods, particularly ReRoPE, demonstrate superior stability in performance across longer code sequences as measured by Edit Sim scores. For instance, ReRoPE consistently achieves the highest Edit Sim scores across all models and languages. This consistent performance indicates that ReRoPE effectively maintains positional coherence throughout extended code sequences, facilitating the capture of syntactical and structural dependencies necessary for code completion tasks. This stability can be attributed to ReRoPE's design, which enables it to adeptly capture and extrapolate positional information by scaling attention scores using a sliding window mechanism.

Conversely, efficient attention mechanisms such as Paged Attention tend to perform poorly in terms of Edit Sim scores. While these methods prioritize optimizing attention computation over long sequences by examining only a portion of the context at any given moment, this efficiency comes at a cost. The limited context can result in a loss of information for tokens that are far apart in the sequence. This is especially detrimental in length extrapolation tasks, where comprehensive context is vital for accurate code completion. As a result, efficient attention mechanisms experience a notable decline in Edit Sim scores in comparison to ReRoPE.

Furthermore, both Flash Attention and StreamingLLM demonstrate significant difficulties in zero-shot code completion tasks, particularly in length extrapolation scenarios, as evidenced by their lower EM and Edit Sim scores. The primary design of these methods prioritizes accelerating attention computation, often at the expense of maintaining critical positional information in extended code sequences. As a result, their performance declines with increasing input length.

\para{Language-Specific Performance}
In context to RQ-3, it is apparent from Table~\ref{tab:result_table} that syntax and structure of different programming languages influence the performance of inference-only methods in length extrapolation scenario. For example, models generally achieve higher Edit Sim scores in Python compared to Csharp and Java. For instance, using Vicuna model, ReRoPE achieves an Edit Sim of $24.630$ in Python, whereas it produces lower score in Csharp ($23.189$) and Java ($21.145$). This finding suggests that Python's more flexible and concise syntax might be easier for models to predict and maintain structural and syntactical integrity over longer sequences. In contrast Csharp and Java, with its strict and detailed syntax, become challenging for LLMs to extrapolate, leading to lower performance. The structured nature of these languages, with its rigid rules and nested formats, makes it harder for the LLMs to maintain high performance when dealing with longer code sequences.

\para{Need of Better Evaluation Metrics}
The Exact Match (EM) metric does not adequately consider minor variations that do not impact the code's functionality or logical correctness. For example, differences in formatting, variable naming, or equivalent syntactical structures are penalized under EM, despite the fact that they can result in functionally identical code. This rigid evaluation approach may reduce the practical utility of the model, especially in real-world situations where preserving functionality is more important than ensuring exact replication of code sequences.

The Edit Similarity (Edit Sim) metric complements EM by leveraging the structural similarity between predicted and ground truth sequences. While it successfully captures minor variations and provides a more comprehensive view of a model's ability to generate coherent outputs, it still falls short in evaluating the functional correctness of the generated code. This is because Edit Sim primarily focuses on syntactical and structural alignment, often disregarding whether the generated code behaves as intended or meets logical task requirements. For instance, a piece of code with high structural similarity to the ground truth but containing a critical logical error would still receive a high Edit Sim score, rendering it insufficient for a comprehensive evaluation.

Given the limitations of existing evaluation metrics such as Edit Sim and EM, there is a pressing need for novel metrics tailored to the unique demands of the long code completion task. These metrics should assess not only the syntactical and structural accuracy of generated code but also its functional correctness and logical equivalence to the ground truth.
To provide a more comprehensive evaluation, metrics that evaluate the following aspects are necessary:
\begin{enumerate}
    \item Compilation and Testability: Examine the ability of generated code to compile successfully and pass predefined test cases.
    \item Functional Correctness: Evaluate the generated code's ability to meet specific functionality criteria, such as correct output or behavior.
    \item Code Quality: Incorporate metrics that assess code readability, maintainability, and adherence to best practices, such as code style, naming conventions, and commenting standards.
\end{enumerate}

By incorporating these enhanced evaluation criteria, a more realistic and comprehensive assessment of model performance can be achieved, ultimately leading to improved long code completion capabilities.

\section{Conclusions}
In this study, we conducted a comprehensive analysis of inference-only methods, such as positional encoding (e.g., ReRoPE) and efficient attention mechanisms (e.g., Paged Attention), specifically for length extrapolation in zero-shot long code completion tasks. Our findings indicate that ReRoPE effectively maintains structural integrity in long sequences, resulting in consistently higher Edit Similarity (Edit Sim) scores. Conversely, while Paged Attention achieves superior Exact Match (EM) scores, it often fails to preserve the overall structure of the code.

Looking ahead, we intend to explore the synergy between Paged Attention and ReRoPE to enhance both EM and Edit Sim scores in zero-shot code completion tasks. We plan to perform experiments using various long code completion datasets and extend this strategy to other code-related challenges to improve the generalizability of our findings. Furthermore, future research could focus on hybrid attention mechanisms and task-specific fine-tuning strategies to bolster the performance of large language models (LLMs) across different programming languages.


The EM metric evaluates model performance by measuring exact matches against the ground truth while disregarding minor variations, such as formatting or variable naming, that do not impact functional correctness. In contrast, Edit Sim captures structural alignment but does not assess the functional correctness of the generated code, potentially allowing logical errors to remain unnoticed. These shortcomings highlight the necessity for new metrics that gauge \textit{functional correctness}, \textit{logical equivalence}, and \textit{code quality}. Such metrics should include criteria like compilation success and test case coverage.

Our work relies on leveraging pretrained LLMs, which presented significant challenges due to computational resource limitations when processing very long context inputs. As a result, we were unable to explore the problem of code completion with extensive contexts. Additionally, we did not investigate the newly released, code-specific LLMs, as they primarily utilize newer versions of the Transformer module, which are not compatible with our current implementation of ReRoPE. Furthermore, the HiRoPE codebase \cite{zhang2024hirope} is unavailable, preventing us from evaluating its performance in long code completion tasks.


\bibliographystyle{ACM-Reference-Format}
\bibliography{ref}
\end{document}